\documentclass[aps]{revtex4}

\def\p {\partial}

\def\be {\begin{equation}}
\def\ee  {\end{equation}}
\def\bea {\begin{eqnarray}}
\def\eea {\end{eqnarray}}
\def\nn {\nonumber}

\begin{document}
\title{Discrete Hamiltonian evolution and quantum gravity}
\author{Viqar Husain$^*$ and Oliver Winkler$^\dagger$}
\affiliation{
$^*$Department of Mathematics and Statistics, University of New Brunswick,\\
Fredericton, NB E3B 5A3, Canada. \\
$^\dagger$Perimeter Institute of Theoretical Physics\\
Waterloo, ON Canada\\
EMail: husain@math.unb.ca, owinkler@perimeterinstitute.ca
}

\thispagestyle{empty}

\date{June 25, 2003}

\begin{abstract}

We study constrained Hamiltonian systems by utilizing general
forms of time discretization. We show that for explicit 
discretizations, the requirement of preserving the canonical Poisson 
bracket under discrete evolution imposes strong conditions on both 
allowable discretizations and Hamiltonians. These conditions permit 
time discretizations for a limited class of Hamiltonians, which 
does not include homogeneous cosmological models. We also present
two general classes of implicit discretizations which preserve Poisson 
brackets for any Hamiltonian. Both types of discretizations generically 
do not preserve first class constraint algebras. Using this observation, 
we show that time discretization provides a complicated time gauge 
fixing for quantum gravity models, which may be compared with the 
alternative procedure of gauge fixing before discretization. 

\end{abstract}

\maketitle

\section{Introduction}

In all approaches to quantum gravity the idea of a discrete
structure of spacetime at the Planck scale is fundamental. In
lattice approaches, such as Regge calculus, dynamical triangulations, 
and other path integral methods, discreteness in space and time is 
built in right from the start\cite{loll}. In canonical methods, such 
as loop quantum gravity, discreteness of space arises as a rather natural
consequence of Dirac quantization \cite{av1,av2,av3,length}, and it 
is possible that a form of discrete time evolution may emerge from 
a regulated action of the Hamiltonian constraint. This is already 
manifested to some extent in spin foam models\cite{foam}, which 
may be viewed as "covariantizations" of loop quantum gravity models.

In attempts to construct a path integral for quantum gravity, a
standard starting point is a classical non-Hamiltonian
discretization using triangulations. It is interesting to ask
whether a phase space path integral for quantum gravity can be defined
starting from a classical discretization that is Hamiltonian. If
possible, this would provide a regularization that is more closely
tied to the classical theory than triangulations or spin foam
methods. There are two main approaches to the problem: (i) fix all
coordinate gauge conditions classically and discretize the reduced
Hamiltonian system, or (ii) attempt a discretization of the
constrained system and study its consequences for the Hamiltonian
structure. The primary requirement for canonical quantization of 
a discrete system is to ensure that the fundamental Poisson bracket 
is preserved under discrete classical evolution. In the second 
approach, the consequences of discretization on the constraint 
algebra also require scrutiny.

These questions have been studied to some extent in the
literature. Discrete evolution schemes for unconstrained
Hamiltonian dynamics appear in \cite{bander}, with 
further refinements in \cite{moncrief}, where schemes preserving 
the Poisson bracket are presented. More recently, the
applicability of discrete evolution schemes to constrained
Hamiltonian theories have been discussed \cite{gp}, with a 
view to applicability to quantum gravity.

From elementary numerical methods, it is known that there are
implicit and explicit approaches to discretization of differential
equations. Although either approach may be useful for studying
classical evolution, explicit methods are more natural for
quantization. This is because discrete Hamiltonians that are
functions of canonical variables at more than one time instant are
effectively non-local in time, so it is unclear whether canonical
quantization makes sense. It is therefore necessary to study
classical discretizations for which the discrete Hamiltonian does
not depend on more than one time point. This may restrict somewhat
the potential applicability of implicit schemes. 

For the purpose of developing a fundamental lattice approach to
quantum gravity starting from a discretization of Hamilton's
equations, it is necessary to ensure that canonical structures are
not affected by the procedures used. This is because classical 
preservation of Poisson brackets is a requirement for a clean 
approach to canonical quantization. The main purpose of this paper is 
to investigate this aspect. We study certain issues associated with 
discrete time evolution. We do not address space discretization, since 
our main concern is the extent to which time discretization can maintain 
Hamiltonian structures for quantization. Specifically we address the 
following questions: (i) What are the general classes of discretization? 
(ii) What are the conditions required to maintain the fundamental 
Poisson brackets under evolution for a given Hamiltonian system? 
(iii) Are there restrictions on allowed Hamiltonians? (iv) How does 
the approach work in simple systems, and (v) How viable is the approach 
for quantum gravity?

Our starting points are general forms of explicit and implicit 
discretizations containing arbitrary functions of phase space variables. 
In Section II we derive the necessary conditions on these functions 
to preserve the fundamental Poisson bracket for both types of schemes. 
We also study the consequences of discretization for constraint evolution
for the parametrized particle. In Section III we apply the approaches 
to the parametrized scalar field, FRW cosmological models, and 
BF theory. In the final section we dicsuss some consequences of our 
results for applicability to quantum gravity.

\section{Discrete Hamiltonian evolution}

As for differential equations, there are two general approaches that 
may be utilised to obtain a discretization of Hamilton's equations that 
preserve the canonical Poisson bracket. The explicit schemes, as the name 
suggests, give explicit functions of the form 
\be 
p_{n+1}= f(q_n,p_n),\ \ \ \ \ q_{n+1}=g(q_n,p_n)
\ee
for evolution. This is in contrast to implicit schemes, where the 
general equations are of the form 
\be 
F(p_{n+1}, q_{n+1},p_n,q_n)=0, \ \ \ \ \ G(p_{n+1}, q_{n+1},p_n,q_n)=0.
\ee
We consider both types of discretization, and outline the conditions 
under which the schemes preserve Poisson brackets, and thus have a 
utility for quantization at the discrete level.

\subsection{Explicit schemes} 

Consider a one particle system with canonical phase space variables 
$(q,p)$,
 and arbitrary Hamiltonian $H(q,p)$.
A first attempt at time discretization of Hamilton's equations
with step $\epsilon$ might be the simplest finite differencing
scheme 
\bea 
q_{n+1} &=& q_n + \epsilon\ {\p H(q_n,p_n)\over \p p_n},\nn \\
p_{n+1} &=& p_n - \epsilon\ {\p H(q_n,p_n)\over \p q_n}
\eea 
However there is an immediate
problem with this: it does not preserve the Poisson bracket under
evolution for even the standard Hamiltonian $H(q,p) = p^2/2m +
V(q)$; the order $\epsilon^2$ term does not vanish unless
$V(q)=0$. To rectify this it is necessary to introduce higher order terms in
$\epsilon$ in the defining equations of the discretization. This
might be done using explicit or implicit finite differencing
schemes. We choose a general explicit approach and study its
consequences. The reasons are simplicity and potential utility for
canonical quantization.  

The defining equations for the discrete evolution we study are the
following:
\bea q_{n+1} &=& q_n + \epsilon\ {\p H(q_n,p_n)\over \p p_n}
+ \epsilon^2\ \alpha(q_n,p_n) \\
p_{n+1} &=& p_n - \epsilon\ {\p H(q_n,p_n)\over \p q_n} +
\epsilon^2 \ \beta(q_n,p_n).
\label{evol1}
\eea
$\alpha,\beta$ are at this stage arbitrary functions. The Poisson 
bracket of the evolved
variables is
\bea
\{ q_{n+1},p_{n+1} \} &=& 1 - \epsilon
\left[ \left\{ q_n,{\p H\over \p q_n} \right\} 
+ \left\{ p_n, {\p H\over \p p_n} \right\} \right] \nn\\
&& + \epsilon^2\
\left[ \{\alpha,p_n\} + \{q_n,\beta\} 
- \left\{ {\p H\over\p p_n},{\p H\over \p q_n} \right\} \right]\nn\\
&& + \epsilon^3\ \left[ \left\{ {\p H\over\p p_n},\beta  \right\} 
- \left\{ \alpha, {\p H\over\p q_n} \right\} \right]\nn\\
&& + \epsilon^4\  \{\alpha,\beta\} \eea
This equation displays the conditions required for preserving the
canonical Poisson bracket. The first order term in $\epsilon$
vanishes identically because the Hamiltonian is a function of the 
discrete
phase space variables at a single time point. (See below for 
discussion of a discretization where this is not the case.) The remaining 
task is to see what functions $H$, $\alpha$, and $\beta$ lead to 
vanishing of the three higher order terms in $\epsilon$. It is 
obviously better to fix $H$ by looking at systems of interest, 
and then attempt solutions for $\alpha$ and $\beta$. 

In order to obtain a general procedure for solving these conditions, 
it is useful to consider first the Hamiltonian independent  
$\epsilon^4$ term. Its solutions can be classified into one of the
following four cases: (i) $\alpha=0$, (ii) $\beta=0$, (iii)
$\alpha(q_n),\beta(q_n)$ or $\alpha(p_n),\beta(p_n)$, and (iv)
$\alpha = \beta\ne 0$. For a given Hamiltonian, the other
conditions can be written out explicitly.

The restrictions that arise may be seen by considering the
standard Hamiltonian
\be
H(q,p)= {p^2\over 2m} + V(q)
\label{hs}
\ee
The conditions have a unique solution: Case (ii) gives
\be \alpha(q) = {1\over m} {\p V(q)\over \p q} \label{a1} \ee
The remaining cases (i),(iii) and (iv) do not give a solution.
This result is surprising given that there are numerous
possibilities for discretizing classical differential equations.
The lesson is that preserving canonical Poisson brackets imposes
rather strong conditions on allowable discretizations.

As a further illustration consider the parametrized particle. This
case is also useful for seeing what happens to the constraints under
discrete evolution. The canonical variables are $(t,p^t)$ and
$(q,p)$, which depend on a time parameter $\lambda$. 
The Hamiltonian constraint is
\be 
H := p^t + {p^2\over 2m} + V(q) = 0 
\label{pp} 
\ee
Following (\ref{evol1}),  discrete evolution with lapse function
$N$ is defined by
\bea q_{n+1} &=& q_n + N\epsilon\  {p_n\over m}
+ \epsilon^2\ \alpha(q_n,p_n) \nn \\
p_{n+1} &=& p_n - N\epsilon\ {\p V(q_n)\over \p q_n} +
\epsilon^2 \ \beta(q_n,p_n).\nn \\
p^t_{n+1} &=& p^t_n \nn \\
t_{n+1} &=& t_n + N\epsilon. \label{dpp} \eea
From these equations it is clear that preservation of the Poisson
bracket goes through in the same way as for the unparametrized
case, with $\beta=0$ and $\alpha$ given by (\ref{a1}) up to a
factor of $N$.

The second check is the evolution of constraints using
(\ref{dpp}). Is $H_{n+1}\equiv
H(t_{n+1},p^t_{n+1},q_{n+1},p_{n+1}) =H_n$? A straightforward
computation using the evolution equations gives
\bea
H_{n+1} &=& p_n^t + {1\over 2m}\ \left[  p_n - 
\epsilon N \p V +\epsilon^2\beta \right]^2  + V(q_{n+1})\nn\\
&=& H_n + {\epsilon^2\over 2m}\ \left[ N^2 (\p V)^2 
+ \beta^2 + 2p_n\beta +2m\alpha \right] + {\cal O}(\epsilon^3), 
\label{hn1}
\eea
where $\p V= \p V(q_n)/\p q_n$. From this it is clear that $H_{n+1}$ does not 
equal $H_{n}$: the Taylor expansion of the potential contains terms 
that cannot be cancelled by any choice of the functions $\alpha,\beta$ 
(which are in any case effectively fixed by the requirement of preserving 
the canonical Poisson bracket). The discretization has made the
constraint second class. 

At this stage there are two possible ways to
proceed: (i) impose the Hamiltonian constraint strongly since it is not 
preserved in time, or (ii) fix the free function $N$ such that the additonal 
terms on the right hand side of (\ref{hn1}) vanish. The latter amounts to 
$N$ becoming a fixed function of $n$, and hence the time step $\epsilon N$
also becomes a function of $n$. Thus, making the evolution consistent via (ii) 
requires a variable time step in this sense, which concomitantly means 
fixing the lagrange multiplier. 

From the theory of constrained Hamiltonian systems it is possible to see that 
these two choices are in fact equivalent: Gauge fixing leads to first class 
constraints becoming second class pairs with the gauge conditions. Since 
we started with a  first class system (the parametrized particle), which became 
second class as a result of time discretization, we can ask what time gauge 
condition has been {\it implicitly} imposed by the discretization that led 
to the Hamiltonian constraint not being preserved in time.

To answer this question let us recall the procedure for the continuous case in 
the canonical theory, where time gauge fixing is accomplished by setting a 
function of the phase space variables to equal the time parameter. Consider a 
phase space function $f(t,p^t,q,p)$ and impose the gauge condition
\be \chi := \lambda - f = 0. \ee
This fixes parameter time $\lambda$. Preserving the gauge
condition under evolution requires
\be {d\chi\over d\lambda} = {\p\chi \over \p\lambda} + \{\chi,
NH\}=  0. \ee
This fixes the lapse as a function of $f$
\be
N =\ {1\over \{ f,H\} } \label{lapse}
\ee

Conversely, fixing $N$ first leads via the above equations to a connection 
with the canonical gauge fixing function $f$. Thus a discretization may 
also be determined by fixing the lapse, which may be viewed as imposing a 
variable time step. This of course, is what is routinely done in numerical 
relativity, with the difference that in that case it is not necessary to 
preserve the canonical Poisson bracket for quantization.

For the discrete evolution we have defined, this procedure may be
used to extract the gauge fixing condition. This is done by
first imposing $H_{n+1}=0$ strongly since it is second class. This
gives an equation for $N$. The gauge fixing function $f$ is then
determined by solving (\ref{lapse}), which in general is rather
messy. For example, for the above discretization of the parametrized
particle, it is not possible to solve for $N$ to all orders in 
$\epsilon$ due to the presence of all orders of $N$ in 
$H_{n+1}$ (\ref{hn1}). (This is because the expansion of $V(q_{n+1})$ 
in powers of $\Delta q_n = q_{n+1} - q_n$ contains powers of $N$.)
Thus only an approximate solution for $N$ is possible, up to a specified 
order in $\epsilon$. 
 
The lessons from this section are two-fold: (i) The conditions on
allowable explicit time discretizations are rather strong if canonical
structures are to be preserved. (ii) For constrained theories with
time reparametrization invariance, time discretization is
equivalent to rather complicated gauge fixing conditions. 

\subsection{Implicit schemes}

There are implicit schemes for time discretizations discussed in 
the literature. We mention two examples below  before introducing 
some generalizations. One of these uses a leap-frog numerical scheme 
to construct a consistent Hamiltonian evolution \cite{moncrief}. 
The other method is implicit and uses a "Hamiltonian" that depends on 
phase space variables at neighbouring time steps \cite{gp}.

The leap-frog discretization scheme for the standard Hamiltonian
$p^2/2m + V(q)$ is \cite{moncrief}
\bea q_{n+1} &=& q_n + \epsilon\ {p_n\over m} - {\epsilon^2\over
2m}{\p V(q_n) \over \p q_n} \nn\\
p_{n+1} &=& p_n - {\epsilon\over 2} \left( {\p V(q_n) \over \p
q_n} + {\p V(q_{n+1}) \over \p q_{n+1}} \right). \eea
This is an implicit scheme since the $V(q_{n+1})$ term in the
$p_{n+1}$ equation is determined by first computing $q_{n+1}$.
Without this term it is, up to factors, the same as the one we
derived above from the general form (\ref{evol1}). It preserves
the canonical Poisson bracket under evolution.

The non-local Hamiltonian scheme \cite{gp} is rather unusual.
Evolution equations are derived from a discrete action, which may 
be viewed as a function of coordinates $q_n$ and velocities $v_n$:
\be S[q_n,v_n]=\sum_n L_n(q_n,v_n) = \sum_n
\left[ v_n(q_{n+1}-q_n)-\epsilon H(v_n,q_n) \right]. 
\ee
This resembles a Hamiltonian form. The conjugate momentum is 
defined by $p_{n+1} := \p L_n/\p q_{n+1} = v_n$. This leads to the
following discrete equations for the canonically conjugate pair
$(q_n,p_n)$ (Eqns. (37,43-44) of Ref. \cite{gp}):
\bea
q_{n+1} &=& q_n + \epsilon\ {\p H(q_n,p_{n+1})\over \p p_{n+1} } \nn \\
p_{n+1} &=& p_n - \epsilon\ {\p H(q_n,p_{n+1})\over \p q_n } 
\label{gp}
\eea
It is in the sense manifested by these equations that the
discretised phase space Hamiltonian is non-local in time.
 
The requirement of preserving the Poisson bracket for more 
general implicit schemes is at first sight more complicated than 
the procedure we followed for the explicit case. The computation
of Poisson brackets requires implicit differentials assuming the 
validity of the implicit function theorem for the discretization. 
For a given discretization this leads to restrictions on the 
form of the Hamiltonian. It therefore appears that a general result 
may be difficult to obtain, and that one is stuck with a case by 
case analysis. 

There is however a general approach for finding a large class of 
implicit schemes that are automatically consistent. This involves 
(i) viewing discrete evolution as a canonical transformation from 
the variables $(q_n,p_n)$ to the variables $(q_{n+1},p_{n+1})$, and 
(ii) requiring that the canonical transformation resemble a 
discretization of Hamilton's equations. This approach gives 
two general classes of implicit discretization starting from 
a given Hamiltonian $H(q,p)$. 

To see how this approach works it is useful to consider each of the 
four classes of generating functions for canonical transformations. 
As a first example consider a generating function of the form
\be 
 I(q_{n+1},p_n) = -q_{n+1} p_n + \epsilon\ F(q_{n+1},p_n) + 
\epsilon^2\ G(q_{n+1},p_n) + \cdots  
\ee
The standard canonical transformation rules give the equations 
\bea 
q_{n+1} &=& q_n + \epsilon\ {\p F\over \p p_n} 
             + \epsilon^2\ {\p G\over \p p_n} +\cdots \nn\\
p_{n+1} &=& p_n - \epsilon\ {\p F\over \p q_{n+1}} 
             - \epsilon^2\ {\p G\over\p q_{n+1}} + \cdots.               
\label{i1}
\eea
The discretization prescription is to fix the functional form of 
$F$ to be that of the given Hamiltonian $H(q_n,p_n)$ with the 
replacement $q_{n}\rightarrow q_{n+1}$, ie set 
\be 
F=H(q_{n+1},p_n).
\ee 
The result is an automatically consistent class of implicit discretizations
where the function $G$ is arbitrary.  The limit $\epsilon \rightarrow 0$ gives
the continuous time Hamilton equations.  
 
The second example of this type is provided by the generating function 
\be 
 I(q_n,p_{n+1}) = -q_n p_{n+1} + \epsilon\ F(q_n,p_{n+1}) + 
\epsilon^2\ G(q_n,p_{n+1}) + \cdots  
\ee
The transformation equations are now 
\bea
q_{n+1} &=& q_n + \epsilon\ {\p F\over \p p_{n+1}} 
             + \epsilon^2\ {\p G\over \p p_{n+1}}+\cdots \nn \\
p_{n+1} &=& p_n - \epsilon\ {\p F\over \p q_n}
             - \epsilon^2\ {\p G\over\p q_n} +\cdots                
\label{i2}
\eea
These again give the continuum Hamilton equations in the 
$\epsilon\rightarrow 0$ limit with $F=H(q_n,p_{n+1})$ 
for arbitrary $G$. (A special case of this scheme arises from the 
action viewpoint discussed in \cite{gp}.)

The remaining two types of transformations with generating functions 
depending on the pairs $(q_n,q_{n+1})$ and $(p_n,p_{n+1})$ do not 
give a form resembling discretized Hamilton equations, so the 
two cases discussed above appear to exhaust this approach for 
generating consistent implicit schemes. 

As for the explicit scheme of the last section, it is readily 
verified that constraints are not preserved generically with these 
two types of implicit discretization. For the 
parametrized particle (\ref{pp}) for example, the evolution equations 
with the first method (\ref{i1}) (without the $\epsilon^2$ terms,  
and with lapse $N$) are 
\bea 
q_{n+1} &=& q_n + N\epsilon\  {p_n\over m}. \nn \\
p_{n+1} &=& p_n - N\epsilon\ {\p V(q_{n+1})\over \p q_{n+1}}. \nn \\
p^t_{n+1} &=& p^t_n \nn \\
t_{n+1} &=& t_n + N\epsilon. 
\label{ppi1} 
\eea
Substitution of these into $H_n= p_n^t + p_n^2/2m + V(q_n)$
shows that $H_{n+1}\ne H_n$.

In summary, we have seen that both the explicit and implicit 
schemes are quite restrictive in their respective settings, although 
with the latter applicable to any Hamiltonian. In particular, the
schemes we have considered are two-step, involving only variables 
at times $n$ and $n+1$. Multi-step general schemes cannot be 
represented as canonical transformations in any simple way 
for the implicit case, and checking of Poisson bracket 
preservation appears to be much more tedious for the explicit 
case. It is also apparent from the simple example of the 
parametrized particle that constraints are not preserved 
under discrete time evolution.

\section{Applications}

It is useful to see whether there are viable implementations of
the explicit and implicit discretization discussed in the last section. 
In the following we consider a few examples of models with Hamiltonian 
constraints to see if there are solutions for the functions $\alpha,\beta$ 
in (\ref{evol1}) that preserve the canonical Poisson bracket. To do this we 
follow the procedure described above. We also examine the time gauge 
fixing conditions introduced by the implicit discretizations (\ref{i1}-\ref{i2}).

\subsection{The parametrized scalar field}
As a first application we look at the parametrized scalar field in
a curved spacetime. Its Hamiltonian is
 \be
 \label{ps}
 H = \int N \left[ p^t + \frac{1}{2\sqrt{q}}\ \Pi^2 + {1\over 2}\ \sqrt{q}\ q^{ab}
 \p_a \phi \p_b \phi + \sqrt{q}\ V(\phi)\right]
 \ee
where the canonical coordinates are $(t,p^t)$ and $(\phi,\Pi)$, and 
$q_{ab}$ is the spatial background metric. The discrete evolution 
equations are
\bea
\phi_{n+1} &=& \phi_n + \epsilon N\ \frac{\Pi}{\sqrt{q}} 
+ \epsilon^2 \alpha(\phi_n, \Pi_n) \nn\\
\Pi_{n+1} &=& \Pi_n + \epsilon\ \left[ \p_a \left( \sqrt{q} q^{ab}
\partial_b \phi N \right) - V' N \sqrt{q} \right] 
+ \epsilon^2 \beta(\phi_n, \Pi_n) \nn\\
t_{n+1} &=& t_n + N \epsilon \nn\\
p^t_{n+1} &=& p^t_n.
\eea
From this one calculates $\{ \phi_{n+1},\Pi_{n+1} \}$ to try 
to find functions $\alpha$ and $\beta$ such that
$\{\phi_{n+1},\Pi_{n+1} \} =1$. It turns out that this is not
possible except for the special case $\phi=\phi(t)$, where it
reduces to the case of the parametrized particle (\ref{pp}).

\subsection{Cosmological models}
We consider two models here, FRW coupled to a scalar field and
deSitter space. These serve to illustrate the main problem with
discretization arising from the Hamiltonian constraint of general
relativity: the kinetic term contains a mixture of gravitational
configuration and momentum variables, which makes it difficult to
find a  discretization  that preserves  the canonical
commutation rules under evolution for the explicit scheme.

For both models we use geometrodynamics phase space variables
$(q_{ab},\tilde{\pi}^{ab})$, and the parametrization
\be q_{ab} = a(t)\ e_{ab} \ \ \ \ \ \ \ \ \ \ \ \ \tilde{\pi}^{ab}
= {1\over 3}\ P(t)\ e^{ab}
\label{cosmo}
\ee
where $e_{ab}$ is the flat Euclidean $3-$metric. (The main
features of the analysis are unchanged for the triad-connection
phase space variables.)

The Hamiltonian constraint for flat FRW minimally coupled to a
scalar field is
\be H(a,p)= - \frac{1}{6}\ a^{1/2} p^2 + {1\over 2}\ a^{-3/2}\ 
  \Pi^2 a^{3/2}\ V(\phi), 
\label{FRW}
\ee
where $V(\phi )$ is a potential for the scalar field $\phi$, and
$\Pi$ is its canonical momentum. For De Sitter space, the potential
is replaced by the cosmological constant $\Lambda$.

The discrete evolution equations obtained using (\ref{evol1}) are
\be a_{n+1} = a_n - \epsilon N\ \frac{1}{3}\ a_n^{1/2} p_n +
\epsilon^2 \alpha(a_n,p_n)
\ee
for both the FRW and deSitter case. The momentum evolution
equation is
\be p_{n+1} = p_n + \epsilon N\ \left[ \frac{1}{12}\ a_n^{-1/2}
p_n^2 +\frac{3}{2}\ a_n^{5/2} \Pi^2 - \frac{3}{2}\ 
a_n^{1/2} V(\phi) \right] + \epsilon^2 \beta(a_n,p_n)
\ee
for the FRW case, and
\be
p_{n+1} = p_n + \epsilon N\ \left[\frac{1}{12} a_n^{-1/2}
p_n^2 +\frac{3}{2}\ a_n^{5/2} \Lambda \right] + \epsilon^2
\beta(a_n,p_n)
\ee
for deSitter. (We ignore the evolution equations for the scalar
field, since these are not relevant for the point we wish to make.) 
The main observation again is that computing  $\{a_{n+1},p_{n+1} \}$ 
shows that there is no choice for $\alpha(a_n,p_n)$ and
$\beta(a_n,p_n)$ which preserves the Poisson bracket to all 
non-vanishing orders in $\epsilon$. 

Thus we conclude that this general form of discretization 
is not applicable to models derived from general relativity.  

\subsection{$BF$ theory}
This example of a fully constrained dynamics is of particular
interest in the context of spin foam models. Once again we focus
on time discretization, which is what is relevant for preserving
the Poisson bracket under evolution.\\

\noindent\underbar{{\it Abelian case}}: the continuum Hamiltonian is given by
\be
 H = \int \left( \lambda \epsilon^{ab} F_{ab} + \mu \partial_a E^a \right) d^3x
\ee
with the two constraints $F_{ab}=0$ and $\partial_a E^a =0$ and
Lagrange multipliers $\lambda$ and $\mu$. The discrete evolution
equations for the canonical variables $A_a$ and $E^a$ are
\bea
A_{a,n+1} &=& A_{a,n} - \epsilon\ \p_a \mu_n, \\
E^a_{n+1} &=& E^a_n + \epsilon\ 2\epsilon^{ab} \p_b \lambda_n. 
\eea
From these it is obvious that the Poisson brackets are trivially
preserved, even without considering terms of order $\epsilon^2$. \\

\noindent \underbar{{\it Non-Abelian case:}} The only difference is that the 
canonical variables are now Lie-algebra valued. Thus the
Hamiltonian is
\be H = \int ( \lambda^i \epsilon^{ab} F^i_{ab} + \mu^i D_a E^{ai}
) d^3x
\ee
where $D_a = \p_a + A_a$ is the covariant derivative. As a result, 
the discrete evolution now becomes non-trivial:
\bea
A^i_{a,n+1} &=& A^i_{a,n} - \epsilon\ \ D_a \mu^i_n + \epsilon^2\ 
\alpha(A^i_{a,n},E^{ai}_n)  \\
E^{ai}_{n+1} &=& E^{ai}_n + \epsilon\ \left[ 2\epsilon^{ab} D_b
\lambda^i_n + \epsilon^{ijk} E^{aj}_n \mu^k_n \right]
   + \epsilon^2\ \beta(A^i_{a,n},E^{ai}_n).
\eea
As a consequence of the additional terms compared to the Abelian
case, it is straightforward to check that there is no  solution 
for $\alpha$ and $\beta$ that leads to a preservation of the Poisson 
bracket.

For the implicit schemes the Poisson bracket is of course preserved 
by the general argument given in the last section. The constraint 
associated with time evolution in BF theory is a combination of 
the two "internal symmetry" constraints, as occurs for example in 2+1 
gravity \cite{bengt}. Thus it is expected that there is some partial 
breaking of these symmetries due to time discretization. Given this, it is 
useful to see what are the (partial) gauge fixing conditions induced by 
the time discretization. 

Denoting the continuum gauge conditions by $f^{i}(A,E)=0$ 
and $g^i(A,E) =0$, the relation between these and the lagrange multipliers 
$\lambda^i$ and $\mu^i$ are given by the requirement that the gauges are 
preserved in time. Schematically this is 
\be 
\{ f, F[\lambda] + G[\mu]\}=0, \ \ \ \ \ \ \ \ \ \{ g, F[\lambda] + G[\mu]\}=0.
\label{fg}
\ee
For the discretization, one can in principle extract the functions 
$\lambda^i$ and $\mu^i$ by setting the time evolved constraints 
to zero, ie. $G^i_{n+1}\equiv G^i(A_{n+1},E_{n+1})=0$ and 
$F^i_{n+1}\equiv F^i(A_{n+1},E_{n+1})=0$. 
One can then attempt to solve Eqn. (\ref{fg}) for the gauge conditions
$f$ and $g$. To get an idea of what the resulting equations look like, the 
first task is to extract $\lambda^i$ and $\mu^i$ from the evolved constraints. 
For the first implicit scheme (\ref{i1}) for example, where $H= H[A_{n+1},E_n]$, 
these equations are 
\bea
G^i_{n+1} &=& G^i 
+ \epsilon\ \left[ - \p \frac{\delta H }{\delta A^i_{n+1}} 
+\epsilon^{ijk}\left( E^{ak}_n \frac{\delta H }{\delta E^j_n} 
- A^j_{n+1} \frac{\delta H }{\delta A^k_{n+1}}\right)\ \right] 
- \epsilon^2\ \epsilon^{ijk} \frac{\delta H }{\delta E^j_n}
\frac{\delta H }{\delta A^k_{n+1}} = 0 \\
F^i_{n+1} &=& F^i_n + \epsilon\ 
\left[\p \frac{\delta H }{\delta E^i_n} 
+ 2\ \epsilon^{ijk}A_n^j \frac{\delta H }{\delta E^k_n}\right] 
+ \epsilon^2\ \epsilon^{ijk} 
\frac{\delta H }{\delta E^j_n}\frac{\delta H }{\delta E^k_n} =0  
\eea
up to order $\epsilon$, where we have suppressed $\epsilon^{ab}$ and 
the spatial indices ($a,b\cdots$). Inserting
\be 
\frac{\delta H }{\delta E^{i}_n} = \epsilon^{ijk} A^j_{n+1} \mu^k_n
\ee
and 
\be  
\frac{\delta H }{\delta A^i_{n+1}} = 2\left(\p \lambda^i_n
      +  \epsilon^{ijk} A^j_{n+1} \lambda^k_n\right)
\ee
into the equations above, one can in principle solve for $\lambda$ and 
$\mu$, and substitute in (\ref{fg}) to get the functions $f$ and $g$. 
These equations give an indication of the rather complex gauge fixing 
induced by this discretization. 

Note that although we are dealing here with time discretization only, this
may still affect the first class nature of the purely internal symmetry
constraints: the first class property is preserved
automatically under evolution if $G^i_{n+1}=0$ follows from $G^i_n=0$, without
fixing any Lagrange multipliers. We have seen above that this is not the case.
Therefore, although time discertization does not affect the internal symmetry
algebra at fixed time ($n$), it nevertheless makes the constraints second class
with respect to evolution in time.
 
From the above examples, the pattern of calculation is clear for application 
to quantum gravity. It is unlikely that explicit schemes will work due to 
non-conservation of the Poisson bracket. The two classes of implicit schemes 
can be pursued to some extent at least, in the connection-triad canonical 
variables. A first attempt might be to keep space continuous and discretize time 
for the general form of the constraints with the Barbero-Immirzi parameter. 
It is possible there are choices of this parameter that simplify the 
form of the constraints and gauge fixing conditions.  

\section{Discussion}

We have studied general classes of explicit and implicit 
time discretizations of Hamilton's equations containing 
arbitrary functions. The general forms were motivated by 
simplicity, and relative ease in establishing the primary 
criterion that the Poisson bracket is preserved under discrete 
evolution. 

The explicit class of schemes we studied preserves 
the Poisson bracket for only a limited class of Hamiltonians.
Therefore its use for quantum gravity is rather limited.
The two classes of implicit schemes have more potential utility 
in that these apply to any Hamiltonian. However the resulting 
gauge conditions are quite complicated. 

Our approach appears to exhaust explicit and implicit two-step 
discretization schemes that preserve the Poisson bracket. In both 
cases, continuum first class constraints are not preserved in time.  
Therefore gauge fixing conditions are implicit in the discretization, 
and can in principle be explicitly recovered. Given this one can 
ask what is the utility of this lattice approach to gauge fixing, 
versus the alternative path of gauge fixing in the continuum theory 
followed by discretization (where preservation of constraints is 
obviously not an issue). For example, the problem of time is "solved" 
by gauge fixing in both approaches. Furthermore, at the level of 
quantization, the old question of whether different gauge fixings 
lead to the same quantum gravity model would reemerge in a lattice 
disguise. 

\bigskip

\noindent{\it Acknowledgements} We thank Thomas Thiemann for discussions. 
This work was supported in part by the Natural Science and Engineering 
Research Council of Canada.


\end{document}